\documentstyle[12pt,axodraw]{article}
\def\lsim{\mathrel{\rlap{\lower4pt\hbox{\hskip1pt$\sim$}}
    \raise1pt\hbox{$<$}}}      
\def\gsim{\mathrel{\rlap{\lower4pt\hbox{\hskip1pt$\sim$}}
    \raise1pt\hbox{$>$}}}      

\begin{document}
\begin{center}
{\Large \bf Spontaneous breaking of the lepton number and  invisible majoron in a 3-3-1 model  } \vskip .3cm \normalsize { C. A de S.
Pires$^a$  and P. S. Rodrigues da Silva$^{a,b}$} \vskip .3cm \it (a)
Departamento de F\'{\i}sica, Universidade Federal da
Para\'{\i}ba, Caixa Postal 5008, 58051-970, Jo\~ao Pessoa - PB,
Brazil.\\
\it (b) Instituto de  F\'{\i}sica Te\'{o}rica, Universidade
Estadual Paulista, Rua Pamplona 145,
01405-900 S\~{a}o Paulo - SP, Brazil.\\
\vskip .3cm
\end{center}
\begin{abstract}
 In this work we implement the spontaneous breaking of the lepton number in the version II of the 3-3-1 models  and study their phenomenological consequences. The main result of this work is that our majoron is invisible
even though it belongs to a triplet representation by the 3-3-1 symmetry.
\end{abstract}
\section{Introduction}
There is already a vast literature concerning the class of models
with gauge structure, $SU(3)_C \times SU(3)_L \times U(1)_N$,
(3-3-1)~\cite{ppf,footpp,early,versionI,versionII}. In these
models the anomaly cancellation requires a minimal of three
families (or a multiple of three in larger versions). Besides,
there is a bunch of new particles and interactions which make
these models phenomenologically rich and attractive as an
alternative to the Standard Model (SM). However if we assume that 
in the realm of intermediate energy there are
no exotic leptons, then the 3-3-1 symmetry allows for only two
possible gauge models for the strong and electroweak
interactions. We refer to such models as  version I and version
II. The version I is the one suggested by Pisano--Pleitez and
Frampton~\cite{ppf}. In this version the standard leptons 
compose the following triplet $(\nu_L \,, l_L \,,l^c_R)^T$. The version II is the
3-3-1 model with right-handed neutrinos~\cite{footpp}. In it the
 standard leptons  constitute the following triplet $(\nu_L \,, l_L \,
\nu^c_R)^T$.

 One peculiar aspects of 3-3-1 models is that 
the Peccei-Quinn~(PQ)~\cite{pq} and the lepton number symmetries emerge
naturally in both versions and their scalar sector provides a
simple implementation of the spontaneous breaking of such
symmetries~\cite{pal,majoronI}.

Despite  the fact that  lepton symmetry
is of great interest for particle physics, since its violation is a necessary  condition to generate Majorana mass term for neutrinos, it was scarcely developed in both versions of the 3-3-1 models~\cite{majoronI}. For example, in what concern  lepton symmetry, an explicit implementation of its spontaneous breaking is missing in the version II of the 3-3-1 models. On the other side, PQ symmetry has received great attention in both versions \cite{alex,app}

In view of this, in this work we implement the spontaneous
breaking of the lepton number symmetry and  discuss some of their consequences in the
version II of the 3-3-1 models. 

This work is organized as follows. In Sec.~(\ref{sec2}) we present
the particle content of the model. Next, in 
Sec.~(\ref{sec3}) we implement the spontaneous breaking of the
lepton number and identify the majoron. In Sec.~(\ref{sec4})
we concentrate on the phenomenology of our majoron. In Sec.~(\ref{sec5})
we  discuss neutrino masses. Finally, in
Sec.~(\ref{sec6}), we present our conclusions.

\section{The model }
\label{sec2}

Our investigation on this work relies on the version II of the
3-3-1 models~\cite{footpp}. Its lepton content comes in the
fundamental representation of the $SU(3)_L$, composing the
following triplet,
\begin{eqnarray}
f_{aL} = \left (
\begin{array}{c}
\nu_a \\
e_a \\
\nu^{c}_a
\end{array}
\right )_L\sim(1\,,\,3\,,\,-1/3)\,,\,\,\,e_{aR}\,\sim(1,1,-1),
 \end{eqnarray}
with $a=1,2,3$ representing the three known generations. We are
indicating the transformation under 3-3-1 after the similarity
sign, ``$\sim$''.

In the quark sector, one generation comes in the triplet
fundamental representation of $SU(3)_L$ and the other two compose
an anti-triplet with the following content,
\begin{eqnarray}
&&Q_{iL} = \left (
\begin{array}{c}
d_{i} \\
-u_{i} \\
d^{\prime}_{i}
\end{array}
\right )_L\sim(3\,,\,\bar{3}\,,\,0)\,,\,\,\,Q_{3L} = \left (
\begin{array}{c}
u_{3} \\
d_{3} \\
u^{\prime}_{3}
\end{array}
\right )_L\sim(3\,,\,3\,,\,1/3),\nonumber \\
&&u_{iR}\,\sim(3,1,2/3)\,,\,d_{iR}\,\sim(3,1,-1/3)\,,\, d^{\prime}_{iR}\,\sim(3,1,-1/3)\nonumber \\
&&u_{3R}\,\sim(3,1,2/3)\,,\,d_{3R}\,\sim(3,1,-1/3)\,,\,u^{\prime}_{3R}\,\sim(3,1,2/3),
\label{quarks} 
\end{eqnarray}
where $a=1,2,3$ and $j=1,2$ both representing the different
generations. The primed quarks are the exotic ones but with the
usual electric charges.

In order to generate the correct mass for all massive particles,
the model requires only  three triplets of scalars, namely,
\begin{eqnarray}
\eta = \left (
\begin{array}{c}
\eta^0 \\
\eta^- \\
\eta^{\prime 0}
\end{array}
\right ),\,\rho = \left (
\begin{array}{c}
\rho^+ \\
\rho^0 \\
\rho^{\prime +}
\end{array}
\right ) ,\, \chi = \left (
\begin{array}{c}
\chi^0 \\
\chi^{-} \\
\chi^{\prime 0}
\end{array}
\right ) , \label{scalarcont} 
\end{eqnarray}
with $\eta$ and $\chi$ both transforming as $(1\,,\,3\,,\,-1/3)$
and $\rho$ transforming as $(1\,,\,3\,,\,2/3)$.

In the gauge sector, the model recovers the standard gauge bosons  and disposes of five more other called  $V^{\pm}$, $U^0$, $U^{0 \dagger}$ and $Z^2$\cite{footpp}. 

Many of these new particles are bileptons (carry two units of lepton number)\cite{app} 
\begin{eqnarray} {\mbox L}(V^+\,,\, U^{\dagger0}\,,\, u^{\prime}_{3} \,,\, \eta^{\prime
0}\,,\,\rho^{\prime +})=-2 \,,\,\,\,\,\,{\mbox L}(V^- \,,\,U^0 \,,\,d^\prime_{i}
\,,\, \chi^ 0\,,\, \chi^-)=+2. \label{leptonnumber} \end{eqnarray}

There are two things that deserve attention in this lepton number distribution above. First, notice that the new quarks, $u^{\prime}_3$ and $d^{\prime}_i$ are
leptoquarks once they carry lepton and baryon numbers. Second, we have two neutral scalars bileptons, $\eta^{\prime
0}$  and $\chi^ 0$. Therefore when one or both of these
neutral scalar bileptons develop a vacuum expectation value (VEV),
we are going to have spontaneous breaking of the lepton number.

In what concerns the potential, it is suitable to note that if we impose lepton number conservation and assume the discrete symmetry $\chi \rightarrow -\chi$, the potential we can form with the three scalar triplets above, 
\begin{eqnarray} V(\eta,\rho,\chi)&=&\mu_\chi^2 \chi^2 +\mu_\eta^2\eta^2
+\mu_\rho^2\rho^2+\lambda_1\chi^4 +\lambda_2\eta^4
+\lambda_3\rho^4+ \nonumber \\
&&\lambda_4(\chi^{\dagger}\chi)(\eta^{\dagger}\eta)
+\lambda_5(\chi^{\dagger}\chi)(\rho^{\dagger}\rho)+\lambda_6
(\eta^{\dagger}\eta)(\rho^{\dagger}\rho)+ \nonumber \\
&&\lambda_7(\chi^{\dagger}\eta)(\eta^{\dagger}\chi)
+\lambda_8(\chi^{\dagger}\rho)(\rho^{\dagger}\chi)+\lambda_9
(\eta^{\dagger}\rho)(\rho^{\dagger}\eta), \label{firstpot} \end{eqnarray}
has the striking feature of providing an extra global symmetry $U(1)$ with the
three triplets of scalars transforming in the following way by
this symmetry:  $\eta\,,\,\rho\,,\, \chi \sim (1)$. The symmetry
can be extended to the entire Lagrangian turning  it a symmetry of
the model\cite{pal}. To accomplish this the multiplets of matter must
transform as $Q_{1L}\sim(1)$, $Q_{il}\sim(-1)$, $f_{aL}\sim(-1/2)$  
and  $e_{aR}\sim(-3/2)$ under the new $U(1)$, with all other multiplets 
not transforming at all. The advantage of having this extra symmetry is that it can
be identified with the PQ symmetry~\cite{pq}, which might
potentially provide a solution to the strong CP problem in the
context of 3-3-1 model~\cite{coment1}. This realization of PQ symmetry in 3-3-1 was first
observed by P. Pal in Ref.~\cite{pal}. Pal also recognized that
such a scenario was not realistic once the spontaneous breaking
of this PQ symmetry implied a Weinberg-Wilczek axion
type~\cite{ww}, already ruled out phenomenologically. What is interesting here is the fact that the PQ symmetry is automatic in the minimal model. 

Despite of the fact that the PQ symmetry in this minimal scenario is useless, it was showed in Ref. \cite{app} that in order to leave the PQ symmetry useful we just have to add a singlet of scalar to the minimal content of the model. Then with the price of introducing a singlet of scalar we have a solution to the strong CP problem. 

In  this paper we will not consider the extension made in Ref. \cite{app} that turned the PQ symmetry a viable symmetry, instead we consider a term in the potential above that breaks explicitly the PQ symmetry, but maintain the lepton number symmetry. For this we have to discard the discrete symmetry $\chi \rightarrow -\chi$ took above. In demanding lepton number conservation, the only possible term that we can add to the potential above is this
\begin{eqnarray} \frac{f}{\sqrt{2}}\epsilon^{ijk}\eta_i \rho_j \chi_k,
\label{tripliceterm} \end{eqnarray}
which explicitly breaks the PQ symmetry. This term is expected to yield a mass for the axion around the scale of
$v_\eta$ which is of order of few hundreds of GeV. 

In summary, the version II of the 3-3-1 models presents two global symmetry, namely, the PQ and the lepton number symmetry. Of these symmetries, only the former was already developed, as  discussed above. The contribution of this work to the development of this version of the 3-3-1 models  is to complete the study of their global symmetries by implementing the spontaneous breaking of the lepton number symmetry.

 The idea in the next section is to break spontaneously the lepton number. As we will see next, we do not need to add anything else to the model in order to implement the  breaking of the lepton number once the scalar content of the model disposes of two neutral scalars bileptons, namely,  $\eta^{\prime o}$  and $\chi^0$. What we have to do is to allow one or both of these  scalars bileptons develop VEV. For sake of simplicity let us develop the case where only $ \eta^{\prime 0} $ develops a VEV. 

\section{Spontaneously broken lepton number }
\label{sec3}
Before we go on, it is important to stress that the model we are 
treating here was built in such way that lepton number is a 
symmetry of the Lagrangian.
The existence of this global symmetry forbids neutrinos of having a 
Majorana mass term. As there is evidence that neutrinos are
massive~\cite{numass} and possibly Majorana-like~\cite{majtype},
there is also a strong motivation to push for the
lepton number violation.  In this section we implement the spontaneous breaking of the lepton number. The case is interesting  because a Goldstone boson appears in the spectrum
which is called majoron. As such majoron comes from a multiplet, it can have
appealing cosmological and astrophysical
consequences~\cite{majastro}. 

 We start expanding
$\eta^{\prime 0}$, $\eta^0$, $\rho^0$  and $\chi^{0 \prime}$ around its VEV, $v_{\eta^{\prime} ,\eta ,\rho ,\chi^{\prime}}$, in the
usual way,
\begin{eqnarray}
\eta^{\prime 0}, \eta^0 , \rho^0 , \chi^{\prime 0} \rightarrow  \frac{1}{\sqrt{2}} (v_{   \eta^{\prime} ,\eta ,\rho ,\chi^{\prime}} 
+R_{\eta^{\prime} , \eta ,\rho ,\chi^{\prime}} +iI_{\eta^{\prime} ,\eta ,\rho ,\chi^{\prime}}). 
\label{vacua} 
\end{eqnarray}
On substituting this expansion in the potential formed with (\ref{firstpot}) and (\ref{tripliceterm}), we obtain the following set of constraints 
\begin{eqnarray} &&\mu^2_\chi +\lambda_1 v^2_{\chi^{\prime}} +
\frac{\lambda_4}{2}(v^2_\eta + v^2_{\eta^{\prime}} ) +
\frac{\lambda_5}{2}v^2_\rho+\frac{\lambda_7}{2}
v^2_{\eta^{\prime}}+\frac{f}{2}\frac{v_\eta v_\rho}
{ v_{\chi^{\prime}}}=0,\nonumber \\
&&\mu^2_\eta +\lambda_2( v^2_{\eta^{\prime}} + v^2_\eta) +
\frac{\lambda_4}{2} v^2_{\chi^{\prime}}
 +\frac{\lambda_6}{2}v^2_\rho +\frac{\lambda_7}{2} v^2_{\chi^{\prime}} =0,
\nonumber \\
&&\mu^2_\eta +\lambda_2( v^2_{\eta^{\prime}} + v^2_\eta) +
\frac{\lambda_4}{2} v^2_{\chi^{\prime}}
 +\frac{\lambda_6}{2}v^2_\rho +\frac{f}{2}
\frac{v_\rho v_{\chi^{\prime}}}{v_\eta} =0,\nonumber \\&&
\mu^2_\rho +\lambda_3 v^2_\rho + \frac{\lambda_5}{2}
v^2_{\chi^{\prime}} +\frac{\lambda_6}{2}( v^2_{\eta^{\prime}}+
v^2_\eta)+\frac{f}{2}\frac{v_\eta v_{\chi^{\prime}}}{v_\rho} =0.
\label{mincondII} \end{eqnarray}
Notice that the second and third constraints imply the relation
\begin{eqnarray} \lambda_7 v^2_{\chi^{\prime}} -f\frac{v_\rho
v_{\chi^{\prime}}}{v_\eta}=0, \label{relation} \end{eqnarray}
which avoids the presence of dangerous tadpoles with
$R_{\chi^{\prime}}$, stemming from the terms
$\lambda_7(\chi^{\dagger}\eta)(\eta^{\dagger} \chi)$ and
$\frac{f}{\sqrt{2}}\epsilon^{ijk}\eta_i \rho_j \chi_k$ in the
potential.

With these  constraints, the potential formed with (\ref{firstpot}) and (\ref{tripliceterm}) leads to the following  mass matrix $M^2_R$ for the neutral
CP-even scalars in the basis $ ( R_\chi \,\, , \,\,
R_{\eta^{\prime}}\,\, , \,\, R_{\chi^{\prime}}\,\, , \,\, R_\eta
\,\, , \,\, R_\rho) $,
\begin{eqnarray}\begin{tiny}\left(\begin{array}{ccccc} -
\frac{\lambda_7v^2_{\eta^{\prime}}}{4} & 0 &
\frac{\lambda_7v_\eta v_{\eta^{\prime}}}{4} & \frac{\lambda_7
v_{\chi^{\prime}} v_{\eta^{\prime}}}{4} &
-\frac{\lambda_7}{4}\frac{v_\eta v_{\chi^{\prime}}
v_{\eta^{\prime}}}{v_\rho} \\
 0 & \lambda_2v^2_{\eta^{\prime}}  & \frac{1}{2}(\lambda_4+\lambda_7)
v_{\chi^{\prime}}v_{\eta^{\prime}} & \lambda_2v_{\eta}
v_{\eta^{\prime}} & \frac{\lambda_6}{2}v_{\eta^{\prime}}v_\rho \\
 \frac{\lambda_7v_\eta v_{\eta^{\prime}}}{4} &
\frac{1}{2}(\lambda_4+\lambda_7)v_{\chi^{\prime}}v_{\eta^{\prime}}
& \lambda_1v^2_{\chi^{\prime}} & \frac{\lambda_4
v_{\chi^{\prime}}v_\eta}{2}+\frac{\lambda_7v_{\chi^{\prime}}v_\eta}{4}
&
 \frac{\lambda_5 v_{\chi^{\prime}}v_\rho}{2}+\frac{\lambda_7}{4}
\frac{v_{\chi^{\prime}}v^2_\eta}{v_\rho} \\
\frac{\lambda_7 v_{\chi^{\prime}}v_{\eta^{\prime}}}{4} &
\lambda_2v_{\eta} v_{\eta^{\prime}}   & \frac{\lambda_4
v_{\chi^{\prime}}v_\eta}{2}+
\frac{\lambda_7v_{\chi^{\prime}}v_\eta}{4}      & \lambda_2
v^2_\eta -\frac{\lambda_7v^2_{\chi^{\prime}}}{4} &
\frac{\lambda_6 v_\eta v_\rho}{2}
+\frac{\lambda_7}{4}\frac{v^2_{\chi^{\prime}}v_\eta}{v_\rho} \\
-\frac{\lambda_7}{4}\frac{v_\eta
v_{\chi^{\prime}}v_{\eta^{\prime}}}{v_\rho} &
\frac{\lambda_6}{2}v_{\eta^{\prime}}v_\rho &  \frac{\lambda_5
v_{\chi^{\prime}}v_\rho}{2}+\frac{\lambda_7}{4}
\frac{v_{\chi^{\prime}}v^2_\eta}{v_\rho} & \frac{\lambda_6 v_\eta
v_\rho}{2}+\frac{\lambda_7}{4}\frac{v^2_{\chi^{\prime}}v_\eta}{v_\rho}
& \lambda_3v^2_\rho-\frac{\lambda_7}{4}
\frac{v^2_{\chi^{\prime}}v^2_\eta}{v^2_\rho}
\end{array}
\right). \label{matrixRII} \end{tiny}\end{eqnarray}
Although the diagonalization of this matrix can be extremely
tough, one can straightforwardly check that it yields a null
eigenvalue by writing the secular equation for its determinant.
This information is all that we need to detect the number of
Goldstone bosons among these real scalars.

For the pseudo-scalars we have the following mass matrix $M^2_I$
in the basis $ ( I_{\eta^{\prime}} \,\, , I_\chi\,\, , \,\,
I_{\chi^{\prime}}\,\, , \,\, I_\eta \,\, , \,\, I_\rho) $
\begin{eqnarray} \left(\begin{array}{ccccc} 
0 & 0 & 0 & 0 & 0 \\
0 & -\frac{
\lambda_7v^2_{\eta^{\prime}}}{4} & \frac{\lambda_7v_\eta
v_{\eta^{\prime}}}{4} & \frac{\lambda_7
v_{\eta^{\prime}}v_{\chi^{\prime}}}{4} & \frac{\lambda_7}{4}
\frac{v_{\chi^{\prime}}v_{\eta^{\prime}}v_\eta}{v_\rho} \\
0 & \frac{\lambda_7 v_\eta v_{\eta^{\prime}}}{4}   & 
-\frac{\lambda_7 v^2_\eta}{4} & -\frac{\lambda_7 v_{\chi^{\prime}}
v_\eta}{4} & -\frac{\lambda_7}{4}\frac{v_{\chi^{\prime}}v^2_\eta}{v_\rho} \\
0 & \frac{\lambda_7 v_{\chi^{\prime}}v_{\eta^{\prime}}}{4} &
-\frac{\lambda_7 v_{\chi^{\prime}}v_\eta}{4} & -\frac{\lambda_7
 v^2_{\chi^{\prime}}}{4} & -\frac{\lambda_7}{4}
\frac{v^2_{\chi^{\prime}}v_\eta}{v_\rho} \\
0 & \frac{\lambda_7}{4}\frac{v_{\chi^{\prime}}v_{\eta^{\prime}}v_\eta}{v_\rho}
 & -\frac{\lambda_7}{4}\frac{v_{\chi^{
\prime}}v^2_\eta}{v_\rho} & -\frac{\lambda_7}{4}\frac{v^2_{\chi^{
\prime}}v_\eta}{v_\rho} &
-\frac{\lambda_7v^2_{\eta}v^2_{\chi^{\prime}}}{4v^2_\rho}
\end{array}
\right). \label{matrixIII} \end{eqnarray}
From this mass matrix we can easily see that the $I_{\eta^{\prime}}$ remains massless and 
decouples from the other pseudo-scalars, $ I_\chi \,\, , \,\,I_{\chi^{\prime}}\,\, , \,\, I_\eta \,\, , \,\, I_\rho$  which, after diagonalization, combine among themselves to generate the Goldstone bosons. These, along with the CP-even Goldstone boson obtained from diagonalization of $M^2_R$, form the set of Goldstone bosons that will be eaten by the massive neutral gauge bosons of the model. We then ended up with an extra massless pseudo-scalar, $J=I_{\eta^{\prime}}$, which
decoupled from the other scalars. This massless pseudo-scalar is the result  of the spontaneous breaking of the lepton number. In literature this pseudo-scalar is the so called majoron. 

It is opportune to remark that, although we have gotten the right number of Goldstones and the majoron, we have not explicitly shown that the true vacuum is the one that we assumed here. As was pointed out some years ago~\cite{minimum}, it is necessary to analyze the possibility of having a broken phase which does not correspond to a minimum of the potential. In other words, we have to be sure that our solution leads to a minimum and not a saddle point. In our case, it would be enough to guarantee that the mass matrices, Eq.~(\ref{matrixRII}) and (\ref{matrixIII}), lead to positive eigenvalues, since they would correspond to positive second derivatives of the potential with respect to the fields at the minimum. We have checked that for values of the $\lambda$'s of order of 0.1 (and $\lambda_7 < 0$) and the VEV's, $v_\eta = v_\rho \approx 100$~GeV, $v_{\eta^\prime}\approx 1$~MeV and $v_\chi\approx 1$~TeV, we obtained that all the mass eigenvalues are positive and above 1~TeV, except for the majoron partner, which has a mass about 3~eV, consistent with what we expect phenomenologically. Then we are safely talking about a true vacuum which corresponds to a minimum of the potential in this case.

Finally, it is important to remember that after the spontaneous breaking of the 3-3-1 symmetry  the triplet $\eta$ in Eq.(\ref{scalarcont}) dissociates into a doublet $( \eta^0\,\,\,\, \eta^-)^T$ plus a singlet $\eta^{\prime 0}$ by the standard $SU(3)_C \times SU(2)_L \times U(1)_Y$(3-2-1) symmetry. As we saw, our majoron comes from $\eta^{\prime 0}$, thus  it is a singlet by the 3-2-1 symmetry. We know singlet majoron in the SM is trivially invisible because it interacts at tree level only with right-handed neutrinos.
We will show in sec.~\ref{sec4} that this is not the case here. Even though our majoron is a singlet by the 3-2-1 symmetry,  we cannot automatically jump to the conclusion that it is an invisible one because it originates from a triplet by the 3-3-1 symmetry, therefore it interacts with the neutral gauge boson  $Z_1$, which play the role of the standard neutral gauge bosons. In this case what threatens the invisibility of our majoron is the decay $Z_1 \rightarrow R_{\eta^{\prime}} +I_{\eta^{\prime}}$. Besides, it 
couples to the charged gauge bosons of the model, $W^{\pm}$ and $V^{\pm}$\cite{footpp}, which leads to an effective coupling to charged leptons  and we have to care about its size, which we do next.
 
\section{The majoron phenomenology }
\label{sec4}

As we saw, the majoron is a massless pseudo-scalar originating
from the spontaneous  breaking of the lepton number global
symmetry. This is possible in extensions of the SM possessing at
least an additional multiplet of scalars. In such extensions, majorons that belong to
either a triplet~\cite{doublet} or a doublet~\cite{triplet} can
interact with neutrinos and charged leptons, being severely
constrained by astrophysical, cosmological~\cite{majastro} and
laboratory data~\cite{concha}. The reason such phenomenological data disfavor usual multiplet majorons is related to the $Z_0$ invisible decay rate~\cite{concha}. The problem comes from astrophysical bounds on the
VEV that breaks the lepton symmetry. Basically, these bounds demand that triplet or
doublet majoron scenarios develop a VEV  around KeV in order to
avoid too fast cooling of red giants~\cite{majastro}. This
constraint is derived through Compton scattering to majoron,
$\gamma + e \rightarrow e + J$. Then, the neutral real scalar
partner of the majoron, let's call it $R_J$, receives a mass of
order of KeV, and therefore contributes to the $Z^0$ invisible decay
$Z^0\rightarrow R_J J$~\cite{concha}. Nonetheless, such a decay is
not allowed for a light $R_J$ since the measured invisible $Z_0$ decay rate is well
explained by three neutrinos and this extra contribution is adding up
to increase this rate by an unacceptable amount. For this reason a successful majoron model is expected to have its origin in a singlet
by the standard 3-2-1 symmetry. A singlet majoron model was first suggested by Chikashige, Mohapatra and Peccei(CMP) in Ref.~\cite{singlet}. In view of this, a criterium to decide if a majoron emerging from  extensions of the SM can be established by demanding that the scalar that gives rise to the majoron be a singlet by the 3-2-1 symmetry. Unfortunately this criterium cannot be used for our majoron.

Despite our majoron be a singlet by the 3-2-1 symmetry, it presents some differences from the CMP majoron. Namely,  our majoron interacts with all the gauge bosons of the model, particularly the $Z^1$, which is the equivalent of the standard neutral gauge boson $Z^0$. Moreover, on evaluating numerically  the matrix $M^2_R$ in (\ref{matrixRII}) for typical values for the parameters involved in it,  we will find that the mass of $R_J$  is proportional to $v_{\eta^{\prime}}$. In face of this, it is recommended that we check if for small $v_{\eta^{\prime}}$ the  decay $Z^1 \rightarrow R_J J$ does not rule out our scenario. 

After the symmetry breaking from 3-3-1 to the $SU(3)_C \times U(1){em}$,  the $Z^0$ get mixed with the $Z^{\prime}$ and form the physical neutral gauge bosons  $Z^1=Z^0 C_\theta -Z^{\prime} S_\theta $ and $Z^2=Z^0 S_\theta +Z^{\prime} C_\theta $\cite{long}. Our interest here lies upon $Z^1$ because it will  play the role of the standard neutral gauge boson\cite{long}. With this mixing we obtain, from the Higgs boson kinetic term $(D_\mu \eta)^{\dagger}(D^\mu \eta)$, the following interaction among the majoron and the neutral gauge boson $Z^1$
\begin{eqnarray}
    {\cal L}_{Z^1R_JJ}=-\frac{1}{3}g\sqrt{3+t^2}S_\theta\left( \partial^\mu R_J J 
    - \partial^\mu J R_J \right) Z^1_\mu 
    \label{Z1-RJ-J}
\end{eqnarray}
where $g$ is the coupling constant for the weak-isospin group $SU(3)_L$ which coincides with the standard one\cite{footpp}, $t=\frac{\sqrt{3}S_W}{\sqrt{3-4S^2_W}}$ and $S_W = \sin \theta_W$ with $\theta_W$ being the Weinberg´s  angle. This interaction leads to the following expression for the decay rate $Z^1 \rightarrow R_J J$
\begin{eqnarray}
    \Gamma_{Z^1 \rightarrow R_J J} = \frac{4}{9}g^2(3+t^2)S^2_\theta m_{Z^1}=\frac{2}{3}S^2_\theta C^2_W (3+t^2)\Gamma_{\nu \nu},
    \label{decayrate}
\end{eqnarray}
where $\Gamma_{\nu \nu}=\frac{G_F m^3_{Z^1}}{12\sqrt{2}\pi}$ is the 
prediction for the decay rate of $Z^1$ into a pair of neutrinos. 
For $S^2_W = 0.23$ which yield  $t=0.57$, we obtain 
\begin{eqnarray}
\Gamma_{Z^1 \rightarrow R_J J}=1.66S^2_\theta \Gamma_{\nu \nu}.
\label{ND}
\end{eqnarray}

From the experimental side, we have \cite{caso}
\begin{eqnarray}
    \Gamma^{ex}_{inv}=(2.993 \pm 0.011)\Gamma_{\nu \nu}
    \label{exp}
\end{eqnarray}
Assuming that there are only three species of neutrinos, 
the window for new physics concerning the invisible decay of $Z^1$ is
\begin{eqnarray}
    \Gamma^{NP}_{inv}\leq 0.004\Gamma_{\nu \nu}.
    \label{NP}
\end{eqnarray}
From Eqs.~(\ref{NP})  and (\ref{ND}) we obtain the constraint 
\begin{eqnarray}
    S_\theta \leq 0.049.
    \label{constraint}
\end{eqnarray}
There is an upper bound on this angle $\theta \leq 0.000132$\cite{long}. As long as this upper bound is obeyed we can safely say that 
the decay $Z^1 \rightarrow R_J J$ 
does not rule out our majoron. 

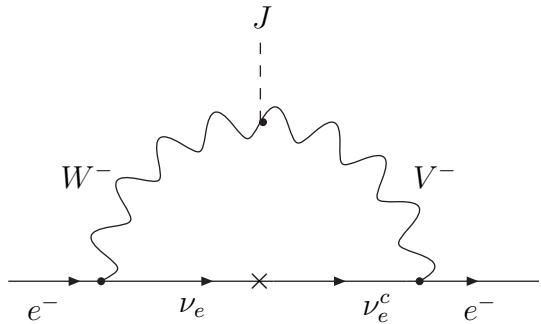
\begin{figure}
\vskip 0.6in \centerline{
\begin{picture}(300,100)(0,0)
\ArrowLine(50,50)(100,50) \ArrowLine(100,50)(150,50)
\ArrowLine(150,50)(200,50) \ArrowLine(200,50)(250,50)
\PhotonArc(145,50)(60,0,180)6 9
\DashLine(145,110)(145,140){4} 
\Text(90,90)[r]{$ W^- $} \Text(220,90)[r]{$V^- $}
\Text(150,150)[r]{$ J $} \Text(150,50)[r]{$ \times $}
\Text(70,40)[r]{$e^-$}  \Text(235,40)[r]{$e^-$}
\Text(125,40)[r]{$\nu_e$} \Text(195,40)[r]{$\nu^c_e$}
\Vertex(85,50){1.5} \Vertex(205,50){1.5} \Vertex(146,110){1.5}
\end{picture}
 } \caption{ Main contribution to $g_{eeJ}$ coupling.}
 \label{fig1}
\end{figure}

Other source of phenomenological constraint against the existence of the majoron arises from its coupling to matter, particularly to electron. The interaction among any lepton and the majoron 
only appears  through loop correction. In regard to electron, the main 
contribution to such interaction is depicted in Fig.~(\ref{fig1}) which originates from the Lagrangian 
\begin{eqnarray}
{\cal L}= \frac{g}{\sqrt{2}}\left( \bar \nu_L \gamma^\mu e^-_L W^+_\mu +\bar \nu^C_L \gamma^\mu e^-_L V^+_\mu	\right) + \frac{g^2}{2}v_\eta W^+ V^- J +\mbox{H.c}.
\label{WVJ}
\end{eqnarray}
With these interactions we obtain the following approximate expression 
for the electron-electron-majoron coupling
\begin{eqnarray}
    g_{eeJ}\simeq \frac{g^4  m_{\nu_e}^2 m_e v_\eta }{16\pi^2 m_W^2 m_V^2}.
    \label{geeJ}
\end{eqnarray}

The experimental constraint is $g_{eeJ}< 10^{-18}$\cite{geeJ}, which is obviously satisfied
by typical values of the parameters involved in Eq.~(\ref{geeJ}). 

Once we are sure that our  majoron is safe from phenomenological constraints, we can go a step further and present what is the main signal of our majoron. According with the Yukawa interactions of this model\cite{footpp}, the triplet $\eta$ only interacts with quarks. It means our majoron does not interact directly with any lepton. Particularly our majoron interactions with fermions involve only leptoquark and ordinary quark\cite{footpp}. In view of this we would expect that 
its main signal be the decay of a leptoquark in ordinary quark plus majoron, 
$q^{\prime} \rightarrow q + J$. The decay rate in this case is
\begin{eqnarray}
    \Gamma(q^{\prime} \rightarrow q + J) = \frac{h^2 m_{q}}{16 \pi}.
    \label{majphe}
\end{eqnarray}
    In this rate $m_q$  stands for the mass of an ordinary quark, while $h$
representing the Yukawa strength interactions among leptoquark-quark-majoron. This means that the discovery of this majoron depends on 
the existence of the exotic quarks $u^{\prime}$  and $d^{\prime}$ which are 
characteristic of the 3-3-1 models.

Let us establish a classification for our majoron. We saw until now that as  consequences of the spontaneous breaking of the lepton number a majoron showed up in the model. In the literature majoron are classified as singlet or multiplet majorons. This classification is based on majorons that come from extensions of the standard model. In view of this, the common majorons are singlet\cite{singlet}, doublet\cite{doublet} or triplet\cite{triplet} majorons. Phenomenological constraints  have ruled out the doublet and the triplet majorons\cite{concha}, allowing for singlet majoron only. We have a peculiar situation here. Our majoron is a triplet by the 3-3-1 symmetry, but it is a singlet by the 3-2-1 symmetry. However it interacts with fermions and gauge bosons, which is typical of multiplet majorons(double or triplet). It is due to this, and to  the fact that our majoron has its origin in a triplet by the 3-3-1 symmetry, that we decided classify our majoron as a multiplet majoron, particularly a triplet majoron. 

We finish this section  pointing out that our majoron presents the peculiar feature of interacting with the quarks instead of interacting with leptons, exactly the contrary to the other multiplet majorons.  It is this fact that turns our majoron phenomenologically distinct from the other multiplet majorons.  

\section{Upper bound on $v_{\eta^{\prime}}$ and neutrino masses } 
\label{sec5}
As the majoron has its origin in a triplet, its associated vacuum $v_{\eta^{\prime}}$ should contribute  to the $\rho$ parameter. In order to check this, let us obtain the expression for $\rho$. The definition of $\rho$  here goes like in the SM: $\rho =\frac{m^2_{W^+}}{m^2_{Z_1}C^2_W}$. The respective expressions for $m^2_{W^+}$  and $m^2_{Z_1}$  in  leading order in  $\frac{1}{v_{\chi^{\prime}}}$ are
\begin{eqnarray}
&&  m^2_{W^+}=\frac{g^2}{4}(v^2_\rho +v^2_\eta -\frac{v^2_\eta v^2_{\eta^{\prime}}}{v^2_{\chi^{\prime}}}) ,\nonumber \\
&& m^2_{Z_1}=\frac{g^2(v^2_\rho +v^2_\eta)}{4 C^2_W}( 1-\frac{3+4t^2}{108}(\frac{5(v^2_\rho +v^2_\eta)}{12v^2_{\chi^{\prime}}}+\frac{9+56\sqrt{2}}{\sqrt{2}}\frac{v^2_{\eta^{\prime}}}{v^2_{\chi^{\prime}}}) ).
\label{massgauge}
\end{eqnarray}
Observe that if we take $v_{\eta^{\prime}}=0$, we recover the masses predicted by the original version of the model\cite{footpp}.

With these masses we obtain the following expression for $\rho$
\begin{eqnarray}
\rho=\frac{v_\rho^2+v_\eta^2 -\frac{v_\eta^2v^2_{\eta^{\prime}}}{v^2_\chi}}{(v_\rho^2+v_\eta^2)( 1-\frac{3+4t^2}{108}(\frac{5(v_\rho^2+v_\eta^2)}
{12v^2_\chi}+\frac{9+56\sqrt{2}}{\sqrt{2}}\frac{v^2_{\eta^{\prime}}}{v^2_\chi}) )}.
    \label{rhoexp}
\end{eqnarray}
The present value for $\rho$  is $\rho =1.0012^{+0.0023}_{-0.0014}$.  Taking appropriate values for the parameters in (\ref{rhoexp}) ( $v_\rho = v_\eta= 10^2$~GeV and  $v_\chi = 10^3$~GeV), we obtain the following upper bound upon $v_{\eta^{\prime}}$
\begin{eqnarray}
    v_{\eta^{\prime}}\leq 40 \,\,\mbox{GeV} .
    \label{upbound}
\end{eqnarray}
The parameter $\rho$  is very sensitive to the value of $v_\chi$. For example, for $v_\chi=500$~GeV we get $v_{\eta^{\prime}}\leq 16 \,\,\,\,\,\mbox{GeV}$.

\begin{figure}
\vskip 0.6in \centerline{
\begin{picture}(300,100)(0,0)
\ArrowLine(50,50)(100,50) \ArrowLine(100,50)(200,50)
\ArrowLine(250,50)(200,50) \DashArrowArc(145,50)(60,0,180)5
\DashLine(145,110)(115,140){4} \DashLine(145,110)(170,140){4}
\Text(90,90)[r]{$ \rho^+ $} \Text(220,90)[r]{$\rho^{\prime+} $}
\Text(115,150)[r]{$ \langle v_\eta \rangle $}
\Text(190,150)[r]{$\langle v_{\eta^{\prime}} \rangle $}
\Text(70,40)[r]{$\nu_a$}  \Text(235,40)[r]{$\nu_a$}
\Text(155,40)[r]{$e^-_a$}
\Vertex(85,50){1.5} \Vertex(205,50){1.5} \Vertex(146,110){1.5}
\end{picture}
 } \caption{ One loop diagram that leads to Majorana neutrino mass.}
 \label{fig2}
\end{figure}
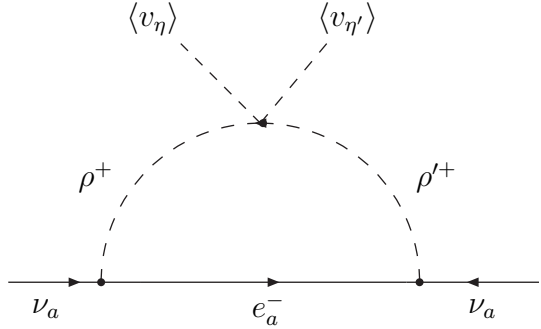

Let us now show that the breaking of the lepton
number through $v_{\eta^{\prime }}\neq 0$  engender Majorana mass for
the neutrinos. One can see this by noticing that when
$\eta^{\prime 0}$ develops a VEV, the Yukawa interaction
$h_{ab}\bar{f}_{aL}e_{bR}\rho$\cite{footpp} together with the term
$\lambda_9(\eta^{\dagger}\rho)(\rho^{\dagger}\eta)$ in the
potential, generate Majorana neutrino mass through one loop as
depicted in Fig.~(\ref{fig2}).
In a naive approximation  such loop provides the following
Majorana mass terms for the neutrinos
\begin{eqnarray}
    m^\nu_{ij} \approx \frac{\lambda_9 h_{ia} m_a h_{aj} v_\eta v_{\eta^{\prime}}}
    {m^2_{\rho^{\prime +}}},
    \label{numass}
\end{eqnarray}
with $i\,\,j=1\,\,2\,\,,3$  and $a=e\,\,,\mu\,\,,\tau$.

On the other side the Yukawa interaction $h_{ab}\bar{f}_{aL}e_{bR}\rho$ generates the following charged lepton mass matrix $m^l_{ab}=h_{ab}v_\rho$. In principle the model has the stage settled to compute masses for all leptons if we can extract information about the matrix elements $h_{ab}$.
This subject deserves careful analysis since only appropriate patterns of mixing could reproduce the recent data on neutrino physics. Although we will study this in detail elsewhere~\cite{progress}, here we can at least check if the matrix elements in (\ref{numass}) can be obtained within a reasonable order of magnitude considering neutrinos acquire masses around eV's. To accomplish this we make some assumptions concerning the values of the parameters involved in lepton masses. Namely, $v_{\rho}\approx 10^2$GeV (responsible for charged lepton masses), $v_{\eta}\approx 10^2$GeV (the Electro-Weak scale), $v_{\eta^{\prime}}\approx 1$MeV (it could be even of the order of KeV's since lepton number is only very softly broken), $m_{\rho^{+ \prime}}\approx 10^3$GeV (since it is a typical scalar related to 3-3-1 symmetry) and $\lambda_{9}\approx 1$. If we take the largest matrix element $h_{ij}$, that related to the tau mass, we see it has to be of order $10^{-2}$, this amounts to
\begin{equation}
 m^\nu_{ij} \approx 10^{-2} \mbox{eV}\,,
\end{equation}
which is an impressive value for the order of magnitude for neutrino mass.
It will be a great achievement to this 3-3-1 model if besides providing an invisible majoron the correct pattern of neutrino mixing and masses emerge naturally from the model.
%

\section{Conclusions }
\label{sec6}
The contribution of this work to the development of the  version II of the 3-3-1 models is the implementation of the spontaneous breaking of the lepton number. The importance of this is in the fact that lepton number violation is a necessary condition to generate Majorana neutrino mass. In view of this the main result of this paper is that spontaneously broken lepton number is viable  once the majoron is invisible. It is important to remember we achieved this without any modification of the minimal model, basically we just allowed $\eta^{ \prime 0}$ to develop a VEV. 

To finalize, in general the 3-3-1 models present two extra global symmetries, namely,  the PQ and the lepton number symmetries. Moreover their scalar sector provides a
simple implementation of the spontaneous breaking of such
symmetries~\cite{pal,majoronI}. Regarding the PQ symmetry, it was shown in Ref.\cite{pal} that in both versions  the spontaneous breaking
of the PQ symmetry imply a Weinberg-Wilczek axion type~\cite{ww} already ruled out phenomenologically,  turning then such symmetry useless. Recently it was showed in Ref.\cite{app} that in the version II  the PQ symmetry regain its usefulness by the addition of a simple scalar singlet. In regard to the lepton number symmetry, it was shown in Ref. \cite{majoronI} that the majoron that comes from the spontaneous breaking of the lepton number in the version I is identical to the Gelmini-Roncadeli one\cite{triplet}, which is already ruled out phenomenologically. In this work we completed this sequence of investigation by showing that in  version II the spontaneous breaking of the lepton number implied an invisible majoron. Such result put the version II in a privileged position. Moreover, as we saw in Sec.\ref{sec5}, it seems that the model has all the ingredients to provide the correct neutrino masses. In this sense, it would be possible to have two strong candidates for cold dark matter and simultaneously solve the neutrino puzzle along with the strong-CP problem~\cite{progress}.

\section*{Acknowledgments}This work was supported by Conselho Nacional de
Pesquisa
e
Desenvolvimento - CNPq (CASP) and Funda\c{c}\~ao de Amparo \'a Pesquisa do Estado 
de S\~ao Paulo - FAPESP (PSRS).



\begin{thebibliography}{99}
\bibitem{coment1}
Notice that there is an
interdependence between PQ symmetry and lepton symmetry in the
sense that the discrete symmetry $\chi \rightarrow -\chi$ alone is
not sufficient to avoid trilinear terms like $ \eta \eta \rho$
and $\chi \chi \rho$ which violate the PQ symmetry. These
trilinear terms are absent only when lepton number conservation
is imposed.
\end{thebibliography}
\end{document}